\documentclass[12pt]{iopart}
\usepackage{iopams}
\usepackage{setstack}

\usepackage{graphicx}
\graphicspath{{images/}}

\usepackage{nameref}
\usepackage{cite}
	
\usepackage{hyperref}
	\hypersetup{
		colorlinks,
		citecolor=black,
		filecolor=black,
		linkcolor=black,
		urlcolor=black
	}

\def\1#1{{\bf #1}}
\def\lp{\left(}
\def\rp{\right)}

\begin{document}

\title[ISCT EoS of Hadrons with Relativistic Excluded Volumes]{Induced Surface and Curvature Tensions Equation of State of Hadrons with Relativistic Excluded Volumes and its Relation to Morphological Thermodynamics}

\author{
	K.A. Bugaev$^{1, 2}$,
	N.S. Yakovenko$^2$,
	P.V. Oliinyk$^3$,
	E.G. Nikonov$^{4, 5}$,
	D. B. Blaschke$^{6,7, 8}$
	L. V. Bravina$^9$
	and
	E. E. Zabrodin$^{9, 10}$
}

\address{$^1$ Bogolyubov Institute for Theoretical Physics, 03680 Kyiv, Ukraine}
\address{$^2$ Department of Physics, Taras Shevchenko National University of Kyiv, 03022 Kyiv, Ukraine}
\address{$^3$ National Technical University of Ukraine "Igor Sikorsky Kyiv Polytechnic Institute", Kyiv, Ukraine}
\address{$^4$ Laboratory for Information Technologies, Joint Institute for Nuclear Research, 141980 Dubna, Russia}
\address{$^5$ Dubna State University, Universitetskaya str. 19, 141980 Dubna, Russia}
\address{$^6$ Institute of Theoretical Physics, University of Wroclaw, Max Born Pl. 9, 50-204 Wroclaw, Poland}
\address{$^7$ Bogoliubov Laboratory of Theoretical Physics, JINR, Joliot-Curie Str. 6, 141980 Dubna, Russia}
\address{$^8$ National Research Nuclear University (MEPhI), Kashirskoe Shosse 31, 115409 Moscow, Russia}
\address{$^9$ Department of Physics, University of Oslo, PB 1048 Blindern,
	N-0316 Oslo, Norway}
\address{$^{10}$ Skobeltsyn Institute of Nuclear Physics, Moscow State University,
	119899 Moscow, Russia}

\begin{abstract}
	An analytical formula that accurately accounts for the Lorentz contraction of the excluded volume of two relativistic hadrons with hard-core repulsion is worked out. Using the obtained expression we heuristically derive the equation of state of Boltzmann particles with relativistic excluded volumes in terms of system pressure and its surface and curvature tension coefficients. The behavior of effective excluded volumes of lightest baryons and mesons is studied at very high temperatures (particle number densities) and for very large values of degeneracy factors. Several parameterizations of the obtained equation of state demonstrate a universal asymptotics of the effective excluded volume at high particle number densities. It is peculiar, that the found maximal packing fraction $\eta \simeq 0.75$ of Lorentz contracted particles is very close to the dense packing limit of classical hard spheres of same radius $\eta_{exc} \approx 0.74$. We show that the developed equation of state is the grand canonical formulation of the morphological thermodynamics approach applied to Lorentz contracted rigid spheres.
\end{abstract}



\section{Introduction}

During last years the concept of morphological thermodynamics became a powerful theoretical tool to study the properties of dense fluids of hard spheres \cite{MorTer1,MorTer2} and hard discs \cite{MorTer3} in condense matter physics. Following the Hadwiger theorem \cite{Hadwiger1957, HadwigerTheorem}, the concept of morphological thermodynamics \cite{MorTer1,MorTer2} assumes that the change of free energy of a convex rigid body ${\cal B}$ immersed into the fluid whose state is away both from the critical point and from wetting and drying transitions can be completely described by four thermodynamic characteristics only: the system pressure $p$, the mean surface tension coefficient $\Sigma$, the mean curvature tension coefficient $K$ and the bending rigidity coefficient $\psi$, i.e. $- \Delta \Omega = p V_{\cal B} + \Sigma S_{\cal B} + K C_{\cal B}+ \psi X_{\cal B}$. Here the quantities $V_{\cal B}$, $S_{\cal B}$, $C_{\cal B}$ and $X_{\cal B}$ are, respectively, the volume of rigid body ${\cal B}$, its surface, mean curvature integrated over the surface of ${\cal B}$ and the mean Gaussian curvature integrated over the surface of ${\cal B}$. Introducing the two local principal curvature radii $R_{c1}$ and $R_{c1}$, one can define the quantities $C_{\cal B}$ and $X_{\cal B}$ as $C_{\cal B} = \int\limits_{\partial {\cal B}} d^2 r \frac{1}{2} \left[\frac{1}{R_{c1}}+ \frac{1}{R_{c2}}\right] $ and $X_{\cal B} = \int\limits_{\partial {\cal B}} d^2 r \frac{1}{R_{c1} R_{c2}} $ (the Euler characteristic).

Unfortunately, this advanced approach is formulated in terms of the canonical ensemble variables, namely the temperature $T$ and particle number density $\rho_p$, and, hence, it cannot be directly applied to modeling the properties of the hadronic phase of quantum chromodynamics (QCD). This is so, since in the statistical mechanics of relativistic particles the conserved quantities are energy, momentum, total angular momentum and fundamental charges (baryonic, electric, strange, etc), while the number of particles is not conserved. Therefore, independently of the morphological thermodynamics its analog in the grand canonical ensemble, known as the induced surface and curvature tensions (ISCT) equation of state (EoS), was developed recently in \cite{Nazar19,QSTAT2019,LFtrans3}. In \cite{Nazar19} the ISCT EoS was developed and successfully applied to the one and two-component mixtures of hard spheres and hard discs with the Boltzmann statistics, while in \cite{QSTAT2019} a similar framework was established for quantum particles with hard-core repulsion. An alternative formulation of the ISCT EoS for classical and quantum particles was recently worked out in \cite{LFtrans3}, which, in contrast to \cite{Nazar19,QSTAT2019} accounts for the statistical fluctuations of the mean particle numbers.

The considered examples of EoS \cite{MorTer1,MorTer2,MorTer3,Nazar19,QSTAT2019,LFtrans3} clearly show us that the morphological thermodynamics approach and its grand canonical analog, i.e. the ISCT EoS, provide rather general approach to study the dense mixtures of hard spheres and hard discs. The question of great importance, however, is whether one can extend this approach in order {\it to take into account the Lorentz contraction of relativistic hard spheres?}

It has to be stressed that there is a strong desire of heavy-ion physics community which is reflected in the multiple and years-long efforts \cite{Kapusta,Zhang,Bugaev_1,Bugaev_2, Indians20} to account for the Lorentz contraction of relativistic hard spheres, since special relativity forbids the existence of rigid bodies. There is an apparent reason behind these multiple efforts to formulate the causal EoS for the particles with hard-core repulsion, i.e. to develop the EoS whose speed of sound does not exceed the speed of light. So far, the only EoS of hard spheres which is proved to be causal in the high-pressure limit was developed in \cite{Bugaev_2}. Moreover, it has a principal difference compared to all other EoS suggested in \cite{Kapusta,Zhang,Indians20}: it is based on the relativistic generalization of cluster and virial expansions \cite{Bugaev_2} suggested by J. Mayer \cite{Mayer} and, hence, it automatically acquires the correct behavior of second virial coefficient both at low and high-pressure limits \cite{Bugaev_2}. However, the EoS expressions derived in \cite{Bugaev_2} are rather complicated even for a single sort of hard spheres and, hence, it is very hard to incorporate them into the hadrons resonance gas model \cite{IST3,X2,X3,X4,X5} based on the truncated ISCT EoS which is adopted to low particle densities.

Since a thorough discussion of the close relations between the morphological thermodynamics approach \cite{MorTer1,MorTer2,MorTer3} and the ISCT EoS can be found in \cite{X5}, here we concentrate on the development of the ISCT EoS for the gas of Boltzmann particles with the Lorentz contraction of their eigenvolumes and, of course, with the corresponding modification of their excluded volumes according to the generalized cluster and virial Mayer expansions for relativistic rigid spheres \cite{Bugaev_1}. Therefore, the major aim of the present work is to obtain the mathematical expressions for the ISCT EoS which are sufficiently simple, reliable and convenient to be incorporated in the near future into the hadron resonance gas model instead of its previous versions reported in \cite{IST3,X2,X3,X4}.

We have to stress the fact, that this is a highly nontrivial task since the excluded volume of two relativistic hard spheres depends not only on the momenta of these particles but also on the angle between their momenta (and the latter dependence is not a simple one!). Therefore, from a formal point of view, the number of different excluded volumes in the studied systems is infinite even for a single sort of particles. In order to achieve our goal, we simplify the present task to work out the ISCT EoS which is causal inside the hadron phase of QCD matter even for very large values of the degeneracy factor of studied particles, up to $3-4 \cdot 10^3$, which has the order magnitude of the total number of spin-isospin states of all know hadrons and hadron resonances listed in the Particle Data Group \cite{PDG} tables.

The work is organized as follows. In Sect. \ref{sec:2} we develop the analytic formula for the excluded volume of two Lorentz
contracted ellipsoids. In Sect. \ref{sec:3} this formula is presented in powers of hard-core radii in a form that is convenient for to get the ISCT EoS. The latter is obtained using the heuristic derivation for low pressures and then it is extrapolated to a high-pressure limit, using the framework of the ISCT concept. Sect. \ref{sec:4} is devoted to the analysis of the effective excluded volume of different particles at very high packing fractions and very high temperatures. Our conclusions and perspectives are presented in Sect. \ref{sec:concl}.

\section{Second virial coefficient of Lorentz contracted rigid spheres} \label{sec:2}

Consider the hard sphere particles of the same mass $m$ that move with relativistic velocities in the rest frame of the medium. Assume that ${\bf r}_i$, ${\bf k}_i$ and ${\bf r}_j$, ${\bf k}_j$ are the coordinates and momenta of the $i$-th and $j$-th Boltzmann particle, respectively. Let ${\bf {\hat r}}_{ij}$ be the unit vector $ {\bf {\hat r}}_{ij} = {\bf r}_{ij}/|{\bf r}_{ij}|$, $ {\bf r}_{ij}= |{\bf r}_i - {\bf r}_j|$. According to \cite{Bugaev_1} for a given set of vectors $\left( {\1 {\hat r}}_{ij} , \1 k_i, \1 k_j \right)$ for the pair of Lorentz contracted rigid spheres of radii $R_1$ and $R_2$ there exists the minimum distance between their centers $r_{ij} ({\bf {\hat r}}_{ij}; {\bf k}_i, {\bf k}_j) = \mbox{ min}|\1 r_{ij}|$. Then one can define the relativistic analog of the hard-core potential $u_{ij}$ via the coordinates ${\bf r}_i,{\bf r}_j$ and momenta ${\bf k}_i, {\bf k}_j$ as
\begin{equation} \label{Eq1n}
	u({\bf r}_{i},{\bf k}_i; {\bf r}_j,{\bf k}_j) %
	\left\{ \begin{array}{rr}
		0\,,      & |\1 r_i - \1 r_j| > r_{ij} \lp
		{\1 {\hat r}}_{ij}; \1 k_i, \1 k_j
		\rp \,,                                      \\
		          &                                  \\
		\infty\,, & |\1 r_i - \1 r_j| \le r_{ij} \lp
		{\1 {\hat r}}_{ij}; \1 k_i, \1 k_j
		\rp \,.
	\end{array} \right.
\end{equation}
Using the definition (\ref{Eq1n}), it was possible in \cite{Bugaev_1} to generalize the usual cluster and virial expansions \cite{Mayer} for this momentum dependent potential. The usual second virial coefficient $a_{2} (T)$ can be defined in terms of the relativistic excluded volume $v (\1 k_1, \1 k_2)$ as follows
\begin{eqnarray}
	a_{2} (T) & = &
	\frac{g^2}{ \rho_t^2}
	\int
	\frac{ d^3 k_{1} d^3 k_{2} }{(2\pi \hbar)^6}
	\,e^{\textstyle - \frac{ E( k_{1}) + E( k_{2}) }{T} }
	\,\, v (\1 k_1, \1 k_2) \,\,,
	\label{Eq2n}
	\\
	\label{Eq3n}
	v(\1 {k}_1, \1 {k}_{2}) & = &
	\frac{1}{2}
	\int d^3{r}_{12}\Theta\left(r_{12}
	({\bf {\hat r}}_{12}; {\bf k}_1, {\bf k}_2)-
	|{\bf r}_{12}|\right),
\end{eqnarray}
where the thermal density $\rho_t(T) $ of particles with the degeneracy factor $g$ and mass $m$ is defined as $\rho_t(T) = \frac{g}{(2\pi \hbar)^3} \int d{\1 k} \exp{\left[- E(k)/T \right]}$. Here $E(k) =\sqrt{k^2+m^2}$ denotes the relativistic energy of the particle
having the 3-momentum $\vec k$. For the sake of simplicity in this work the derivations are made for a general case, while the numeric analysis is made only for the particles of the same mass.

As shown in \cite{Bugaev_2}, the Van der Waals (VdW) extrapolation which is causal in the high-pressure limit should be formulated not in terms of the second virial coefficient, but exclusively in terms of $v(\1 {k}_1, \1 {k}_{2})$ which is the relativistic excluded volume. The obtained EoS is free of the causality paradox since at high pressures the main contribution to the momentum integrals corresponds to the smallest values of the relativistic excluded volume \cite{Bugaev_2}. In other words, the smallest values of $v (\1 k_1,\1 k_2)$ are reached, when both spheres are ultrarelativistic, their velocities are collinear and such configurations, as argued in \cite{Bugaev_2}, correspond to the dense packing of the gas of Lorentz contracted hard spheres. Hence we would like to employ these important findings for the ISCT EoS of the Lorentz contracted hard spheres. For this purpose, in the present work, we use the following general expression derived for the parameter $a=1$ in \cite{Bugaev_2} for the ultrarelativistic limit, i.e. for two thin disks
\begin{eqnarray}
	\fl 2v(\1 {k}_1, \1 {k}_{2}) \equiv 2v^{Urel}_{kl} (\Theta_v) = \frac{4 }{3 } \pi \frac{ R_k}{ \gamma_k} \left( R_k +
	R_l \cos^2 \left( \frac{\Theta_v}{2} \right) \right)^2 +
	\nonumber \\
	+ \frac{4}{3} \pi \frac{R_l}{\gamma_l} \left( R_l + R_k \cos^2 \left( \frac{\Theta_v}{2} \right) \right)^2 + 2 \pi a R_k R_l (R_k + R_l) \biggl| \sin \left( \Theta_v \right) \biggr|
	\,, \qquad
	\label{EqI}
\end{eqnarray}
Here $R_l$ is the hard core radius of a spherical particle of sort $l$ in its rest system, $\gamma_l = \sqrt{m_l^2+\vec k_l^2}/m_l$ is the relativistic $\gamma$-factor of a particle with mass $m_l$, which has $3$-momentum $\vec k_l$ in the rest system of the thermostat and the azimuthal angle $\Theta_v$ is the angle between the $3$-momenta of particles $k$ and $l$, and, as in \cite{Bugaev_2}, the axis $OZ$ in the momentum space is directed along the $3$-momentum of a particle of type $k$. However, it should be noted that (\ref{EqI}) is applicable only for the angles $ - \frac{\pi}{2} \le \Theta_v \le \frac{\pi}{2}$ (for the angles $ \frac{\pi}{2} \le \Theta_v \le {\pi}$ one has to replace $ \Theta_v \rightarrow \pi - \Theta_v $). Therefore, in practice, it is more convenient to performe an integration over the spherical angle $\Theta_v$ over the interval $\Theta_v \in [0; \frac{\pi}{2}]$:
\begin{equation}\label{EqIIIa}
	\int\limits_{0}^\pi d \Theta_v \, \sin(\Theta_v)... = 2 \int\limits_{0}^{\pi/2} d \Theta_v \, \sin(\Theta_v) ... \,.
\end{equation}

In order to provide the correct value of second virial coefficient $a_2(T)$ given by (\ref{Eq2n}) one has to integrate the expression (\ref{EqI}) over two spherical angles $\Theta_v$ and $\phi_v$ in momentum space to calculate the average excluded volume
\begin{eqnarray} \label{Eq6n}
	\langle v^{Urel}_{kl} \rangle = \frac{\int\limits_0^\pi d \Theta_v \sin(\Theta_v) \int\limits_0^{2\pi} d \phi_v \,
		v^{Urel}_{kl} (\Theta_v)}{\int\limits_0^\pi d \Theta_v \sin(\Theta_v) \int\limits_0^{2\pi} d \phi_v} = \frac{1}{2} \int\limits_0^\pi d \Theta_v \sin(\Theta_v) v^{Urel}_{kl} (\Theta_v)
	\,.
\end{eqnarray}
Right the averaged excluded volume (\ref{Eq6n}) can be compared to the traditional excluded volume for the non-relativistic particles of different shapes. Therefore, in the present section, the main object of our analysis is $\langle v^{Urel}_{kl} \rangle$. Choosing the value of parameter $\tilde a = \frac{22}{9 \pi} \simeq 0.778$ in (\ref{EqI}), one can exactly reproduce the excluded volume of two non-relativistic hard spheres as follows
\begin{equation}
	2 \langle v^{Urel}_{12}(R_k, R_l) \rangle \Biggl|_{\gamma_l=\gamma_k =1} = \,2v_{12}^{Nrel} (R_k, R_l) \equiv \frac{4}{3} \pi (R_k+R_l)^3 \, ,
	\label{EqIIa}
\end{equation}
\begin{figure}[th!]
	\includegraphics[width=0.5\textwidth]{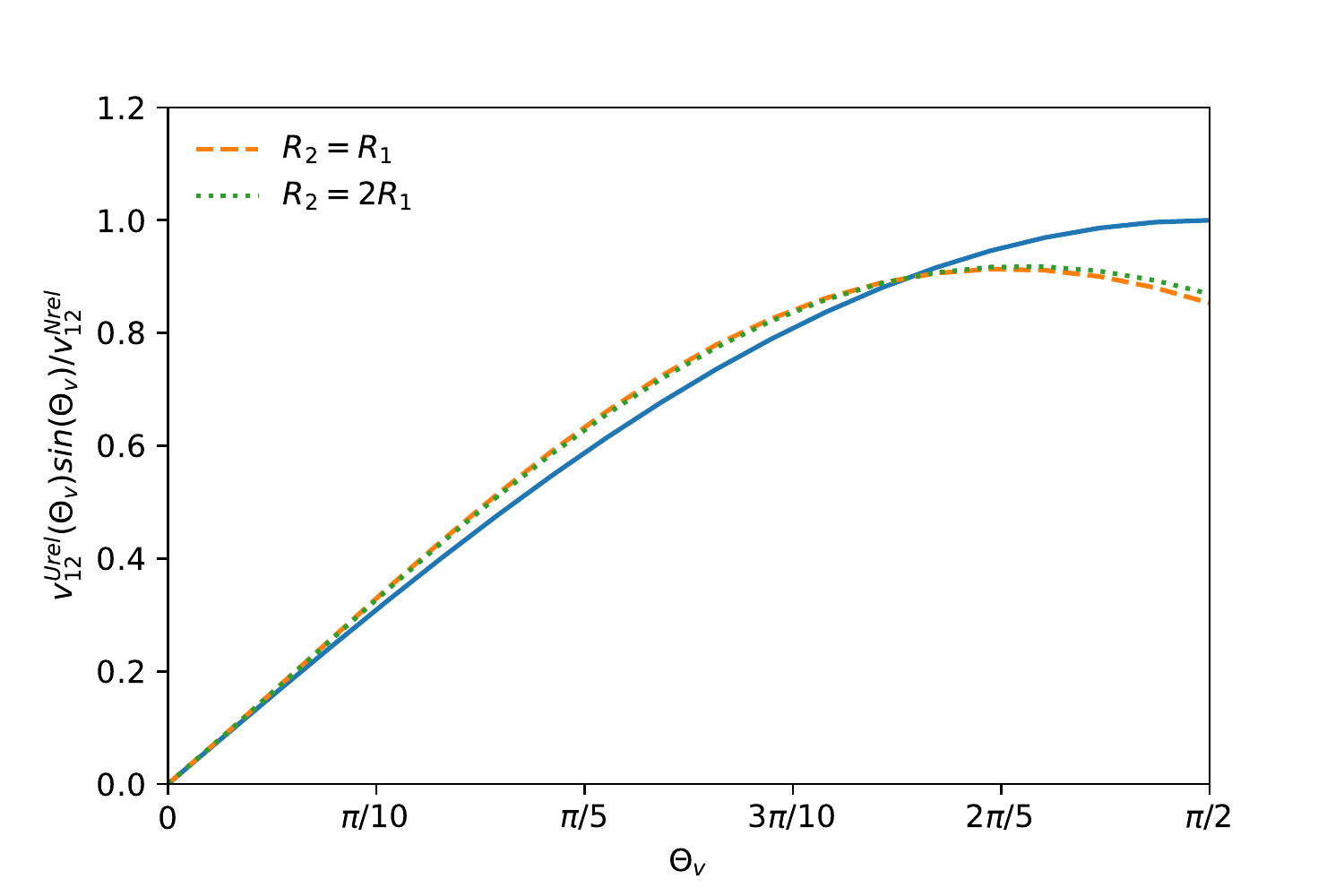}
	\includegraphics[width=0.5\textwidth]{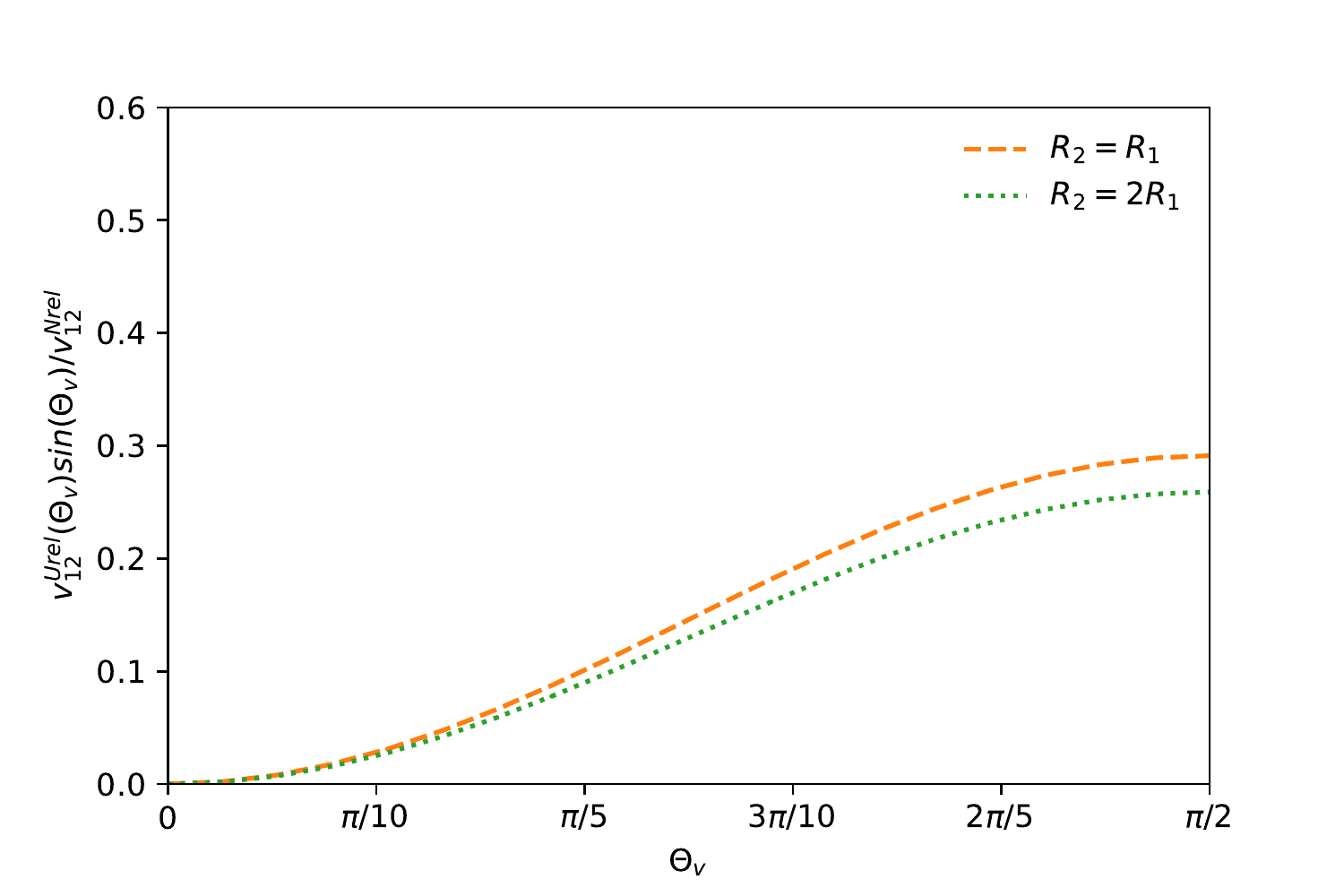}
	\caption{The excluded volume $v^{Urel}_{12} (\Theta_v) \sin(\Theta_v) $ of two Lorentz contracted hard spheres in units of the excluded volume of two nonrelativistic hard spheres $v_{12}^{Nrel}$ for the radii $R_2=R_1$ fm (dashed curve) and for the radii $R_2=2 R_1$ fm (dotted curve) as a function of the angle $\Theta_v$ between the $3$-momentum vectors of the particles. \textit{Left panel} shows the nonrelativistic limit $\gamma_1=\gamma_2 = 1$ (for two spheres). The solid curve is the exact result for $v_{12}^{Nrel} \sin(\Theta_v) $, while the dashed and dotted curves are obtained from (\ref{EqI}). \textit{Right panel} shows the ultra-relativistic limit $\gamma_1=\gamma_2 = 1000$ (for two thin disks). }
	\label{fig1}
\end{figure}
As shown in the left panel of figure \ref{fig1}, the approximate result for two non-relativistic spheres of equal radii obtained from (\ref{EqI}) differs from the exact one by $5 - 10\%$ on a narrow interval of spherical angle values $\Theta_v \in [\frac{2\pi}{5}; \frac{\pi}{2}]$. But this difference is compensated by a slight difference in the interval of angles $\Theta_v \in [\frac{\pi}{10}; \frac{3\pi}{10}]$. As a result, the areas under the curves are equal. The dotted curve in the left panel of figure \ref{fig1} demonstrates that (\ref{EqI}) for the radii $R_2=2 R_1$ also correctly describes the excluded volume of two non-relativistic hard spheres.

On the other hand, the reduction of the coefficient $a$ value from $a=1$ to $\tilde a\simeq 0.778$ does not affect the accuracy of (\ref{EqI}) in the limit of two ultra-relativistic hard spheres, since for high densities, the last term on the right-hand side of (\ref{EqI}) does not contribute to the system pressure, because for high pressures the configurations with $\Theta_v \neq 0$ are strongly suppressed. The reason for such a suppression is clearly visible from figure \ref{fig1}. In the ultra-relativistic limit for two particles $\gamma_1 = \gamma_2 \gg 1$ the relative excluded volume $ v^{Urel}_{12} (\Theta_v) /v_{12}^{Nrel} $ is extremely small for the collinear arrangement of $3$-momenta of the particles, while it is finite in the case of their perpendicular arrangement. At high pressures $p$ the probability of state with the excluded volume $ v^{Urel}_{12} (\Theta_v)$ of pair of particles 1 and 2 is about $\exp\left[ - p \, v^{Urel}_{12} (\Theta_v) /T \right] \rightarrow 0$, i.e. it is vanishing for $ v^{Urel}_{12} (\Theta_v)/v_{12}^{Nrel} > 0$.

Thus, the coefficient $\tilde a\simeq 0.778$ does not lead to a modification of the EoS in the ultrarelativistic limit, but allows one to reproduce the nonrelativistic limit more accurately compared to other approximate formulas. Let us estimate now the deviation of expression (\ref{EqI}) from the exact value for the case of an infinitely thin disk and a sphere with radiuses $R_d$ and $R_s$, respectively, i.e. for the case when one sphere is ultra-relativistic and the other is non-relativistic. To simplify the calculations, we assume that $R_k=R_d=R_s$. Then the exact value $v_{sd} = \frac{10+3\pi}{3} \pi R_s^3 = \frac{10+3\pi}{32} \cdot 2v_{12}^{Nrel} (R_s, R_s) \simeq 0.607 \cdot 2v_{12}^{Nrel} (R_s, R_s)$, while (\ref{EqI}) gives one
\begin{eqnarray}\label{Eq9n}
	&&2 \langle v^{Urel}_{12} \left(R_1=R_s, R_2=R_s \right) \rangle \Biggl|_{\gamma_1=1, \gamma_2=\infty}
	\simeq 0.6137 \cdot 2v_{12}^{Nrel} \left(R_s, R_s\right)
	\,,
\end{eqnarray}
which differs by only $1.16\%$ from the exact result. In general case for a pair of the non-relativistic and the ultra-relativistic spheres the double excluded volume is
\begin{equation}\label{Eq10n}
	2v_{sd}\left(R_s, R_d\right) = \pi \left[2 R_s \left(R_s^2+R_d^2\right) - \frac{2}{3}R_s^3 + \pi R_s^2 R_d \right]
	\,.
\end{equation}
Table \ref{tab:1} shows that the averaged excluded volume $\langle v^{Urel}_{12} \left(R_1=R_s, R_2=R_d\right) \rangle \Biggl|_{ ^{\gamma_1=1}_{\gamma_2=\infty} }$ found for this case from (\ref{EqI}), differs from the exact value $v_{sd}\left(R_s, R_d\right)$ by about $1\%$ only. This means that compared to (\ref{EqI}) with $a=1$ for two ultra-relativistic particles the value of $\tilde a\simeq 0.778$ coefficient significantly improves the accuracy of the approximate formula (\ref{EqI}) in this intermediate case as well. Hence, (\ref{EqI}) with the parameter $\tilde a\simeq 0.778$ can be safely used for arbitrary values of gamma-factors of particles.

\begin{table}[t!]
	\begin{center}
		\begin{tabular}{c|c|c|c}
			$R_d$         & $ v_{s d}/v_{12}^{Nrel} (R_s, R_d)$ & $ \langle v_{12}^{Urel} \rangle / v_{12}^{Nrel} (R_s, R_d)$ & rel. dev. (\%) \\ \hline \hline
			$R_d=R_s$     & 0.6065                              & 0.6137                                                      & 1.16           \\
			$R_d=0.5 R_s$ & 0.7559                              & 0.7643                                                      & 1.10           \\
			$R_d=2 R_s$   & 0.4334                              & 0.4377                                                      & 0.98           \\
		\end{tabular}
	\end{center}
	\caption{Comparison of the averaged excluded volume $\langle v^{Urel}_{12} ( R_s, R_d) \rangle $ (third column) with the exact results (second column) found for the non-relativistic hard sphere of radius $R_s$ and the ultra-relativistic one (hard-disc) of the radius $R_d$. A few choices of $R_d$ are indicated in the first column, while the relative deviation of the result obtained with (\ref{EqI}) is shown in the last column.}
	\label{tab:1}
\end{table}

\section{Derivation of ISCT EoS for the relativistic excluded volumes} \label{sec:3}

The relativistic excluded volume formula (\ref{EqI}) that was thoroughly discussed in the preceding section is the starting point for obtaining the ISCT EoS. Our next step is to obtain such EoS in a heuristic way. Introducing the eigen volume $v_j$, eigen surface $s_j$ and eigen perimeter $c_j$ of particles of type $j$, one can rewrite (\ref{EqI}) in the form
\begin{eqnarray}
	\label{Eq11n}
	\fl 2\hat v^{Urel}_{kl} (\Theta_v) = \hat v_k + \hat s_k( \Theta_v) R_l + \hat c_k ( \Theta_v) R_l^2 + \hat v_l +\hat s_l( \Theta_v) R_k + \hat c_l ( \Theta_v) R_k^2
	\,, \quad \hat v_k \equiv \frac{v_k}{\gamma_k}
	\,, \\
	\label{Eq11b}
	\fl \hat s_k( \Theta_v) \equiv s_k \left[ \frac{1}{2\gamma_k}\left(1+\frac{\Delta_{c1}}{2} \right) + \frac{a}{2} \biggl| \sin \left( \Theta_v \right) \biggr| \right]
	, ~
	\hat c_k ( \Theta_v) \equiv c_k \frac{1}{2\gamma_k}\left(1+\frac{\Delta_{c1}}{2}+ \frac{\Delta_{c2}}{6} \right)
	\,,~
	\\
	\label{Eq12b}
	\fl
	v_k = \frac{4}{3}\pi R_k^3
	\,, ~\, s_k =4 \pi R_k^2
	\,, ~\, c_k = 4 \pi R_k
	\,, ~\, \Delta_{c1} \equiv \cos(\Theta_v) - 1
	\,, ~\, \Delta_{c2} \equiv \cos^2(\Theta_v) - 1
	\,,
\end{eqnarray}
which is convenient for the subsequent analysis. Note that the choice of the coefficients $ \hat s_k( \Theta_v)$ and $ \hat c_k ( \Theta_v) $ is not unique in general. The present choice corresponds to the work on the derivation of ISCT EoS for classical hard spheres and hard discs \cite{Nazar19} and it allows us to use the parameterizations obtained in \cite{Nazar19}.

Since (\ref{Eq11n}) for $2\hat v^{Urel}_{kl} (\Theta_v)$ is symmetric over the indexes $k$ and $l$, it does not matter what particle to choose for integrating over the angle $\Theta_v$ in symmetric terms. Therefore, the two leading terms of the virial expansion for the grand canonical ensemble of the Lorentz contracted hard spheres can be written in the following form \cite{Bugaev_2}:
\begin{eqnarray}
	\label{Eq15n}
	&& p = T \sum_k \hat \phi_k e^{\frac{\mu_k}{T}} - T \sum_k \sum_l \hat \phi_k e^{\frac{\mu_k}{T}} \hat \phi_l e^{\frac{\mu_l}{T}} \hat v^{Urel}_{kl}
	\,,\\
	\label{Eq16n}
	&& \hat \phi_l (T) = 4 \pi g_l \int \frac{d k_l\, k_l^2}{(2 \pi \hbar)^3} e^{- \frac{\sqrt{m_l^2 + k_l^2}}{T}} \int\limits_0^\frac{\pi}{2} d\Theta_l \sin(\Theta_l)
	\,,
\end{eqnarray}
where $p$ is the pressure of Boltzmann gas of particles with temperature $T$ and a set of chemical potentials $\{\mu_l\}$, and $\hat \phi_l (T)$ is an analog of the thermal density of gas of particles with mass $m_l$ and spin-isospin degeneracy factor $g_l$. The summations in (\ref{Eq15n}) are performed among all sorts of particles and their antiparticles are considered as independent constituents. It is important to keep in mind that the hat over the thermal density (\ref{Eq16n}) means that the double integration over momentum and angle acts as an operator on any function with the same hat appearing to the right of the function $\hat \phi_l (T)$.

Here we generalize and employ the approaches proposed in \cite{Bugaev_2,X2,Sagun14} for extrapolating the pressure (\ref{Eq15n}) to high densities. First, we substitute expression (\ref{Eq11n}) into the formula (\ref{Eq15n}) and collect similar terms. As a result, for low densities we obtain the pressure in the following form:
\begin{eqnarray}
	\label{Eq17n}
	\fl p =T \sum_k \hat \phi_k e^{\frac{\mu_k}{T}} - T \sum_k \sum_l \hat \phi_k e^{\frac{\mu_k}{T}} \hat \phi_l e^{\frac{\mu_l}{T}} \frac{1}{2} \left[ \hat v_k + \hat s_k( \Theta_v) R_l \right.
		\nonumber \\
		+ \left. \hat c_k ( \Theta_v) R_l^2 + \hat v_l +\hat s_l( \Theta_v) R_k + \hat c_l ( \Theta_v) R_k^2\right] =
	\\
	\fl = T \sum_k \hat \phi_k e^{\frac{\mu_k}{T}} \left[ 1 - \hat v_k \sum_l \hat \phi_l e^{\frac{\mu_l}{T}} + \hat s_k( \Theta_v) \sum_l \hat \phi_l R_l e^{\frac{\mu_l}{T}} + \hat c_k ( \Theta_v) \sum_l \hat \phi_l
		R^2_l e^{\frac{\mu_l}{T}} \right] \simeq
	\nonumber \\
	\label{Eq18}
	\fl \simeq T \sum_k \hat \phi_k \exp \left[ \frac{\mu_k}{T} - \hat v_k \sum_l \hat \phi_l e^{\frac{\mu_l}{T}} - \hat s_k( \Theta_v) \sum_l \hat \phi_l R_l e^{\frac{\mu_l}{T}} - \hat c_k ( \Theta_v) \sum_l \hat \phi_l R^2_l e^{\frac{\mu_l}{T}} \right]
	\,,
\end{eqnarray}
where to derive (\ref{Eq18}), first, we doubled the identical terms in the double sum on the right-hand side of (\ref{Eq17n}) and, second, moved the small terms into the exponential function. Such an approximation is, evidently, valid for low densities (low pressures), but, in contrast to the usual VdW extrapolation, such approximation led to a causal EoS in the high-pressure limit \cite{Bugaev_2}. Therefore, we exploit it here as well.

For low densities we use the approximation for the total pressure $p \simeq T \sum_l \phi_l e^{\frac{\mu_l}{T}}$ and introduce the new variables $ \Sigma = T \sum_l \phi_l R_l e^{\frac{\mu_l}{T}}$ and $K = T \sum_l \phi_l R^2_l e^{\frac{\mu_l}{T}}$. Next we rewrite the right hand side of (\ref{Eq18}) in terms of functions $ \Sigma$ and $ K$ and postulate the following equation for pressure $p$
\begin{eqnarray}
	\label{Eq19n}
	&&\frac{p}{T} = \sum_k \hat\phi_k \exp{\left[ \frac{\mu_k}{T} - \hat v_k \frac{p}{T} - \hat s_k \frac{ \Sigma}{T} - \hat c_k \frac{ K}{T} \right]}
	\,.
\end{eqnarray}
For the functions $\Sigma$ and $K$ it is reasonable to require that their definitions should contain the distribution functions which are similar to the ones used in the definition of pressure $p$. Hence, we postulate the following system
\begin{eqnarray}
	\label{Eq20n}
	&&\frac{ \Sigma}{T} = \sum_k \hat\phi_k R_k \exp{ \left[ \frac{\mu_k}{T} - \hat v_k \frac{p}{T} - \hat s_k \frac{\Sigma}{T} - \hat c_k \frac{ K}{T} \right]}
	\,,\\
	\label{Eq21n}
	&&\frac{ K}{T} = \sum_k \hat\phi_k R^2_k \exp{\left[ \frac{\mu_k}{T} - \hat v_k \frac{p}{T} - \hat s_k
			\frac{ \Sigma}{T} - \hat c_k \frac{ K}{T} \right]}
	\,.
\end{eqnarray}
to be valid for all temperatures and particle number densities.

The above treatment is a typical extrapolation of the low-density EoS to high densities, which is similar to the usual VdW extrapolation. It is necessary to stress that, despite the complicated velocity dependence, the form of the $k$-th particle-free energy $- f_k = \hat v_k {p} + \hat s_k {\Sigma} + \hat c_k {K}$, which depends only on the system pressure $p$, surface tension coefficient ${\Sigma}$ and curvature tension coefficient $K$, corresponds to the expectation of the morphological thermodynamics \cite{Roth1, Roth2}. According to the latter, a decrease of full free energy $- \Delta \Omega$ of a rigid convex body ${\cal B}$ immersed into the fluid contains only contributions linear in the volume $V_{\cal B}$, surface $S_{\cal B}$, integrated mean $C_{\cal B}$ and Gaussian $X_{\cal B}$ curvatures of a body $- f_{\cal B} = V_{\cal B} p + S_{\cal B} \sigma + C_{\cal B} K + X_{\cal B} \psi$. Here $\sigma$, $K$ and $\psi$ denote surface, curvature and Gaussian curvature tension coefficients, respectively. Considering the particle of $k$-th sort as a rigid body ${\cal B}$, we obtain the same result with one exception that the quantity $\psi$ does not appear in the system (\ref{Eq19n}) - (\ref{Eq21n}).
Of course, one can write a separate equation for $\psi$, but it seems that this is not necessary, since the whole term $\bar X_k \psi$ has a dimension of energy and, hence, it only redefines the value of chemical potential $\mu_k$ of $k$-th sort of particles, which in the grand canonical ensemble is an independent variable. This qualitative derivation of the ISCT EoS for the Lorentz contracted hard spheres based on the postulates of morphological thermodynamics provides additional arguments for justifying the system (\ref{Eq19n}) - (\ref{Eq21n}).

As shown in \cite{X2,Nazar19}, the system (\ref{Eq19n}) - (\ref{Eq21n}) can be generalized further in order to go beyond the usual VdW approximation by introducing the additional adjustable parameters $ \{\alpha_k, \beta_k, k=1, 2, ... L \}$ and making the following replacements
\numparts
\begin{eqnarray}\label{Eq22n}
	&& R_k \longrightarrow \alpha_p R_k \quad \mbox{in (\ref{Eq20n})}\,, \quad R_k^2 \longrightarrow \beta_p R_k^2 \quad \mbox{in (\ref{Eq21n})}\,,\\
	&& \hat s_k{\Sigma} \longrightarrow \alpha_k^S \hat s_k \Sigma \quad \mbox{and} \quad \hat c_k K \longrightarrow \beta_k^S \hat c_k K \quad \mbox{in (\ref{Eq20n})} \,,\\
	\label{Eq23n}
	&& \hat s_k{\Sigma} \longrightarrow \alpha_k^C \hat s_k \Sigma \quad \mbox{and} \quad \hat c_k K \longrightarrow \beta_k^C \hat c_k K \quad \mbox{in (\ref{Eq21n})} \,,
\end{eqnarray}
\endnumparts
with the coefficients $\alpha_k^S\,, \beta_k^S\,, \alpha_k^C\,, \beta_k^C > 1$ as it will be shown below. Then we arrive at the system of equations
\begin{eqnarray}
	\label{Eq24n}
	&&p \equiv T \sum_k \hat\phi_k \exp{\left[ \frac{\mu_k}{T} - \hat v_k \frac{p}{T} - \hat s_k \frac{\Sigma}{T} - \hat c_k \frac{ K}{T} \right]}
	\,, \\
	\label{Eq25n}
	&&\Sigma \equiv \alpha_p T \sum_k \hat\phi_k R_k \, \exp{ \left[ \frac{\mu_k}{T} - \hat v_k \frac{p}{T} - \hat s_k \alpha_k^S \frac{ \Sigma}{T} - \hat c_k \beta_k^S \frac{K}{T} \right]}
	\,, \\
	\label{Eq26n}
	&& K \equiv \beta_p T \sum_k \hat\phi_k R^2_k \, \exp{\left[ \frac{\mu_k}{T} - \hat v_k \frac{p}{T} - \hat s_k \alpha_k^C \frac{ \Sigma}{T} - \hat c_k \beta_k^C \frac{ K}{T} \right]}
	\,, \qquad
\end{eqnarray}
which is assumed to be valid for arbitrary densities. It is clear that this system is equivalent to the original equation (\ref{Eq15n}) at low densities, since all the modifications in the quantities $\Sigma$ and $K$ due to the introduction of additional terms which are proportional to $\hat v_k$, $ \hat s_k $ and $\hat c_k $ in the exponential functions will appear in the virial expansion only in the terms with the third and higher powers of density \cite{Nazar19, X2}.

One can readily check that for constant values of the functions $\hat v_k = v_k$, $\hat s_k = \frac{1}{2} s_k$ and $ \hat c_k = \frac{1}{2} c_k$ one automatically recovers the ISCT EoS for the classical hard spheres \cite{Nazar19}. Therefore, the system (\ref{Eq24n})-(\ref{Eq26n}) is the ISCT EoS for the Lorentz contracted hard spheres.

As shown in several examples in \cite{Nazar19}, the correct choice of the parameters $\alpha_p, \beta_p, \{\alpha_k^C=\alpha_k^S, \beta_k^S=1, \beta_k^C, k=1, 2, ... L \}$ allows one to reproduce not only the second but also up to the fifth virial coefficients of the gas of non-relativistic hard spheres and even the ones of hard discs with several hard-core radii. Therefore, it seems that the right strategy to employ the system (\ref{Eq24n})-(\ref{Eq26n}) is to determine the parameters $\alpha_p, \beta_p, \{\alpha_k^S, \alpha_k^C, \beta_k^S, \beta_k^C, k=1, 2, ... L \}$ for the non-relativistic hard spheres, first, and then to apply them to the relativistic systems.

From (\ref{Eq24n})-(\ref{Eq26n}) it is seen that the functions $\hat v_k $, $\hat s_k$ and $ \hat c_k$ depend on the spherical angle in momentum space $\Theta_v$. As shortly discussed above such dependence is of crucial importance to reproduce the non-relativistic virial expansion at low densities. Therefore, the attempt \cite{Indians20} to introduce the Lorentz contraction in the IST EoS by the naive replacements of the eigenvolume $v_k$ and eigensurface $s_k$ of particle of sort $k$ by $v_k \rightarrow \gamma_k^{-1} v_k$ and $s_k \rightarrow \gamma_k^{-1}s_k$ is highly unrealistic.

To demonstrate the importance of the coefficients $\alpha_p, \beta_p, \{\alpha_k^C=\alpha_k^S > 1, \beta_k^S=1, \beta_k^C > 1, k=1, 2, ... L \}$ in the ISCT EoS (\ref{Eq24n})-(\ref{Eq26n}) it is instructive to introduce the effective excluded volume of the $k$-th sort of particles and define it via the free energy $f_k$ as
\begin{equation}\label{Eq37n}
	- f_k = \hat v_k p + \hat s_k \Sigma + \hat c_k K \equiv {\hat v_k^{eff} p} \quad \Rightarrow \quad \hat v_k^{eff} = \frac{\hat v_k p + \hat s_k \Sigma + \hat c_k K}{p}
	\,.
\end{equation}
From (\ref{Eq37n}) it is evident that $ v_k^{eff}$ strongly depends on the particle momentum, the spherical angle $\Theta_k$, and system pressure. For further analysis, it is necessary to represent the ratios $\Sigma/p$ and $K/p$ in terms of the partial quantities
$p_l$, $\Sigma_l$ and $K_l$ as
\begin{eqnarray}\label{Eq38n}
	&&\frac{\Sigma}{p} \equiv \alpha_p { \sum\limits_{l = 1}^{L} R_l \, \hat p_l \, \exp\left[ -(\alpha_l^S -1) \hat s_l \frac{\Sigma}{T} - (\beta_l^S -1) \hat c_l \frac{K}{T} \right] } \cdot \left[ \sum\limits_{l = 1}^{L} p_l \right]^{-1}
	\,, \\
	\label{Eq39n}
	&&\frac{K}{p} \equiv \beta_p { \sum\limits_{l = 1}^{L} R_l^2 \, \hat p_l \, \exp\left[ -(\alpha_l^C -1) \hat s_l \frac{\Sigma}{T} - (\beta_l^C -1) \hat c_l \frac{K}{T} \right] } \cdot \left[ \sum\limits_{l = 1}^{L} p_l \right]^{-1}
	\,,
\end{eqnarray}
obtained from (\ref{Eq24n})-(\ref{Eq26n}). Note that in (\ref{Eq38n}) and (\ref{Eq39n}) the partial pressures $ \hat p_l$ acts only on the variables $\hat s_l$ and $\hat c_l$ and do not act on the quantities $p_l$, $\Sigma$ and $K$.

From (\ref{Eq38n}) and (\ref{Eq39n}) one can show that all non-vanishing effective excluded volumes $\{\hat v_l^{eff} \neq 0, l=1, 2, ..., L\}$ gradually decrease, if the system pressure increases. Indeed, (\ref{Eq38n}) and (\ref{Eq39n}) show that for low densities (or low pressures), i.e. for $\Sigma \cdot \max\{\hat s_l\}/T \ll 1$ and $K \cdot \max\{ \hat c_l\}/T \ll 1$, each exponential in these equations can be approximated as $\exp \left[- (\alpha_l^S -1)\hat s_l \Sigma/T \right] \simeq \exp \left[- (\alpha_l^C -1)\hat s_l \Sigma/T\right] \simeq 1$ and $\exp\left[ - (\beta_l^S -1) \hat c_l K /T \right] \simeq \exp\left[ - (\beta_l^C -1) \hat c_l K /T \right] \simeq 1 $. By construction under such inequalities one recovers the usual multi-component VdW result for low densities \cite{Nazar19}. In other words, when the terms with the parameters $\alpha$ and $\beta$ are not important, the system (\ref{Eq24n})-(\ref{Eq26n}) is equivalent to the multi-component VdW EoS given by (\ref{Eq19n})-(\ref{Eq22n}).

In the limit of high densities (or high pressures) it is easy to show the validity of the inequalities $\Sigma {\hat s_l}/T \gg 1$ for any $\hat s_l > 0$ and $K {\hat c_l}/T \gg 1$ for any $\hat c_l > 0$. Indeed, under the conditions $\alpha_l^S > 1$, $\alpha_l^C > 1$, $\beta_l^S > 1$ and $\beta_l^C > 1$ from (\ref{Eq38n}) and (\ref{Eq39n}) one can see that the ratios $\Sigma/p$ and $K/p$ vanish in this limit. Consequently, in the limit of high pressure the effective excluded volume of $k$-th sort of particles $\hat v_k^{eff}$ should be essentially reduced compared to the low pressure limit. From the explicit form of operator $\hat s_l$ (\ref{Eq11b}) it is clear, however, that even for ultra-relativistic particles the configurations with the angles around the values $\Theta \simeq 0$ and $\Theta \simeq \pi$ for which the surface operator vanishes, i.e. $\hat s_l = 0$, can also contribute into $\hat v_k^{eff}$. Moreover, under the choice $\beta_l^S = 1$, the suppression of the induced surface tension coefficient $\Sigma$ will be rather weak compared to the non-relativistic formulation. The quantitative analysis of the effective excluded volume $\hat v_k^{eff}$ for the choice $\beta_l^S = 1$ is given below, since it leads to new interesting results.

\section{Effective excluded volume calculations} \label{sec:4}

To demonstrate the important role of induced surface and curvature tensions, we consider the effective excluded volume $\hat v_k^{eff}$ that is averaged over momenta and angles
\begin{equation}
	\label{EqVIp}
	\langle v_k^{eff} \rangle = \frac{ (\hat \phi_k \, \hat v_k) p + (\hat \phi \, \hat s_k) \Sigma + (\hat \phi \, \hat c_k) K}{ p (\hat \phi_k \, \hat 1)}
	\,.
\end{equation}
In what follows ISCT EoS (\ref{Eq24n})-(\ref{Eq26n}) was applied to the gases of baryons (nucleons and antinucleons) and pions in order to obtain a temperature $T$ dependence of such an effective excluded volume for a number of equations of the state represented with different sets of ISCT EoS parameters (see Table \ref{tab:2}).

\begin{table}[h!]
	\begin{center}
		\begin{tabular}{c|c|c|c|c}
			      & $\alpha_p$ & $\beta_p$ & $\alpha_k^S=\alpha_k^C $ & $\beta_k^C$ \\ \hline \hline
			VdW   & 1          & 1         & 1                        & 1           \\
			IST   & 2          & 0         & 1.245                    & 0           \\
			ISCT  & 1.14       & 1.52      & 1.07                     & 3.76        \\
			ISCT2 & 1.38       & 0.62      & 1.14                     & 3.37        \\
		\end{tabular}
	\end{center}
	\caption{Various sets of parameters of ISCT EoS. The IST set is obtained in \cite{Ivanytskyi18} which reproduces the third and fourth virial coefficients of the gas of non-relativistic hard-core spheres. The ISCT set provides the best description of the Carnahan-Starling EoS of hard spheres \cite{CSeos} for packing fraction $\eta = \in [0; 0.45]$ and the ISCT2 exactly reproduces its virial expansion up to the fifth coefficient according to \cite{Nazar19}. Here $\beta_k^S=1$ was fixed.}
	\label{tab:2}
\end{table}

Note that the VdW set with the coefficients $\alpha_p=\beta_p=1, \{\alpha_k^C=\alpha_k^S=1, \beta_k^S=\beta_k^C = 1, k=1, 2, ... L \}$ formally corresponds to the case of the VdW equation of state with hard-core repulsion and relativistic effects. The IST set with $\beta_p = 0$ takes into account only the contribution of a surface tension coefficient, but not the one of an induced curvature tension coefficient
to the free energy of the system. The ISCT and ISCT2 sets allow for the best description of the whole gaseous phase of a hard spheres gas \cite{Nazar19}, i.e. up to the packing fractions $\eta \equiv \sum_k \rho_k V_k \simeq 0.45$ (here $\rho_k$ denotes the particle number density of the $k$-th sort of particles). Moreover, to use the sets of parameters found previously in \cite{Nazar19} we fix $\beta_1^S=1$.

\begin{figure}[thb!]
	\includegraphics[width=0.5\textwidth]{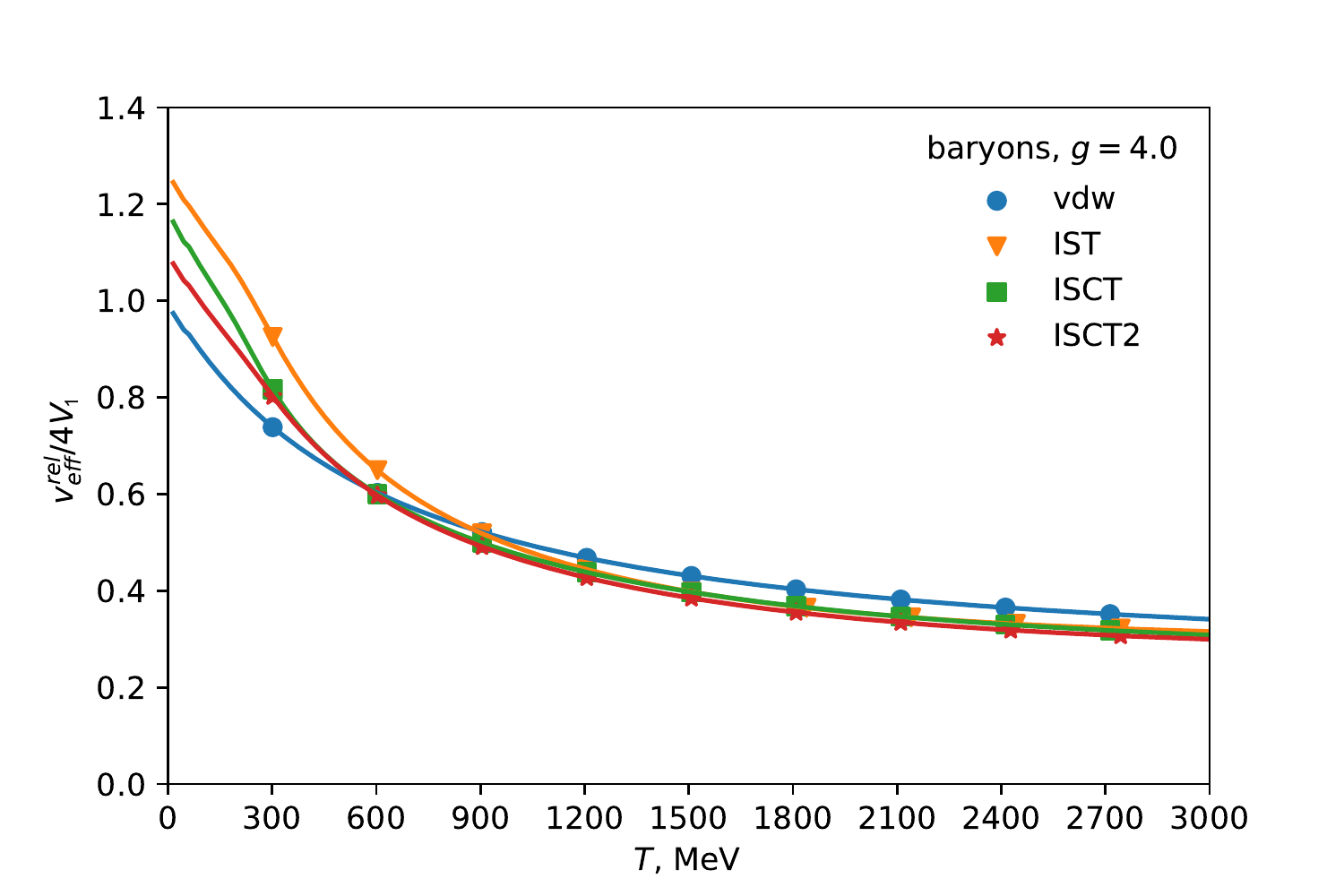}
	\includegraphics[width=0.5\textwidth]{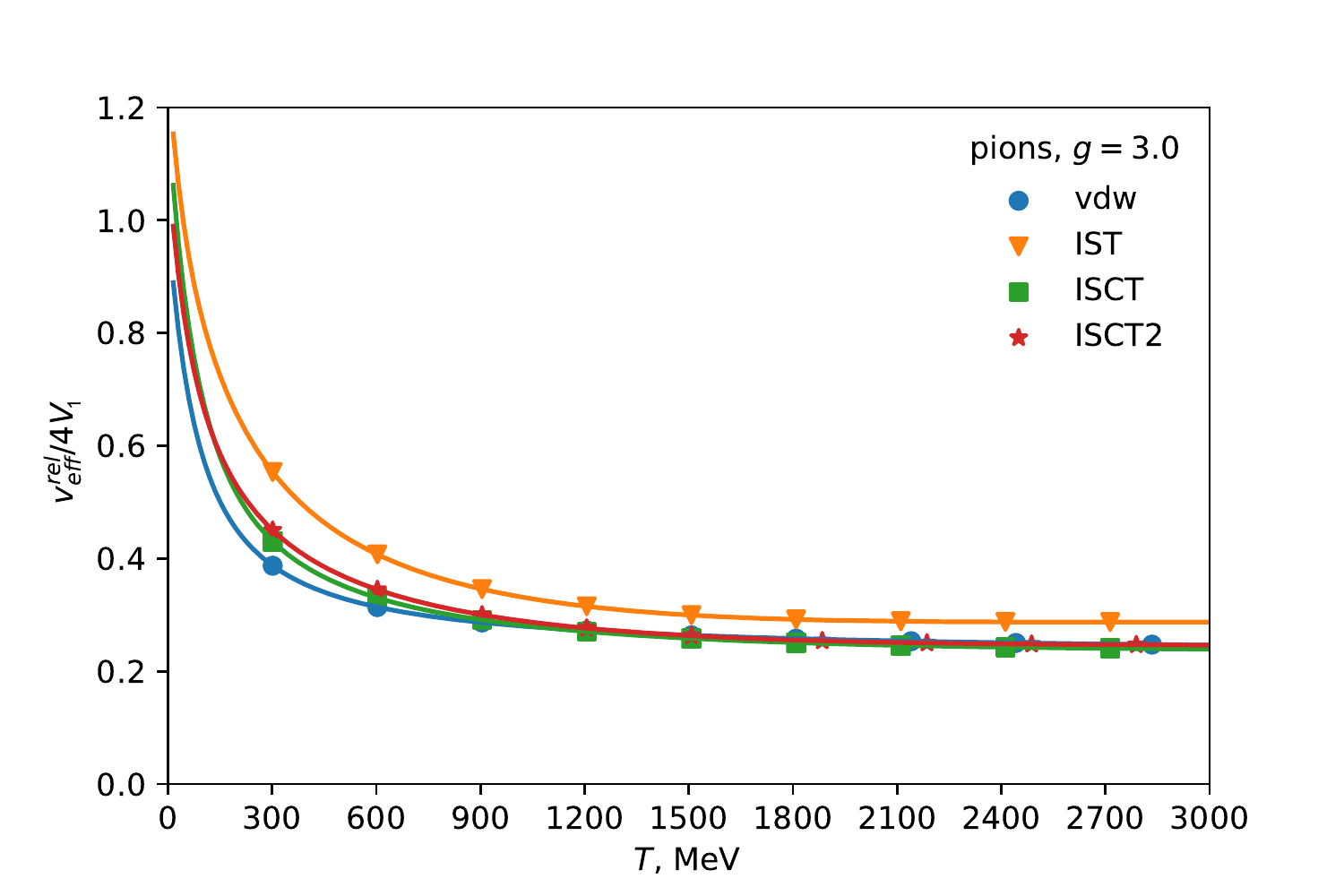}
	\caption{ Averaged excluded volume of gas of Lorentz contracted rigid spheres in units of the excluded volume of two non-relativistic hard spheres of radius $R_1=0.39$ fm as a function of the temperature of the system $T$. Various markers correspond to different sets of ISCT EoS parameters (Table \ref{tab:2}). \textit{Left panel:} a gas of baryons (nucleons and anti-nucleons) with masses $m_1=940$ MeV and degeneracy factor $g_1=4$. \textit{Right panel:} a gas of pions with masses $m_1=140$ MeV and degeneracy factor $g_1=3$.}
	\label{fig:vol1}
\end{figure}

\begin{figure}[thb!]
	\includegraphics[width=0.5\textwidth]{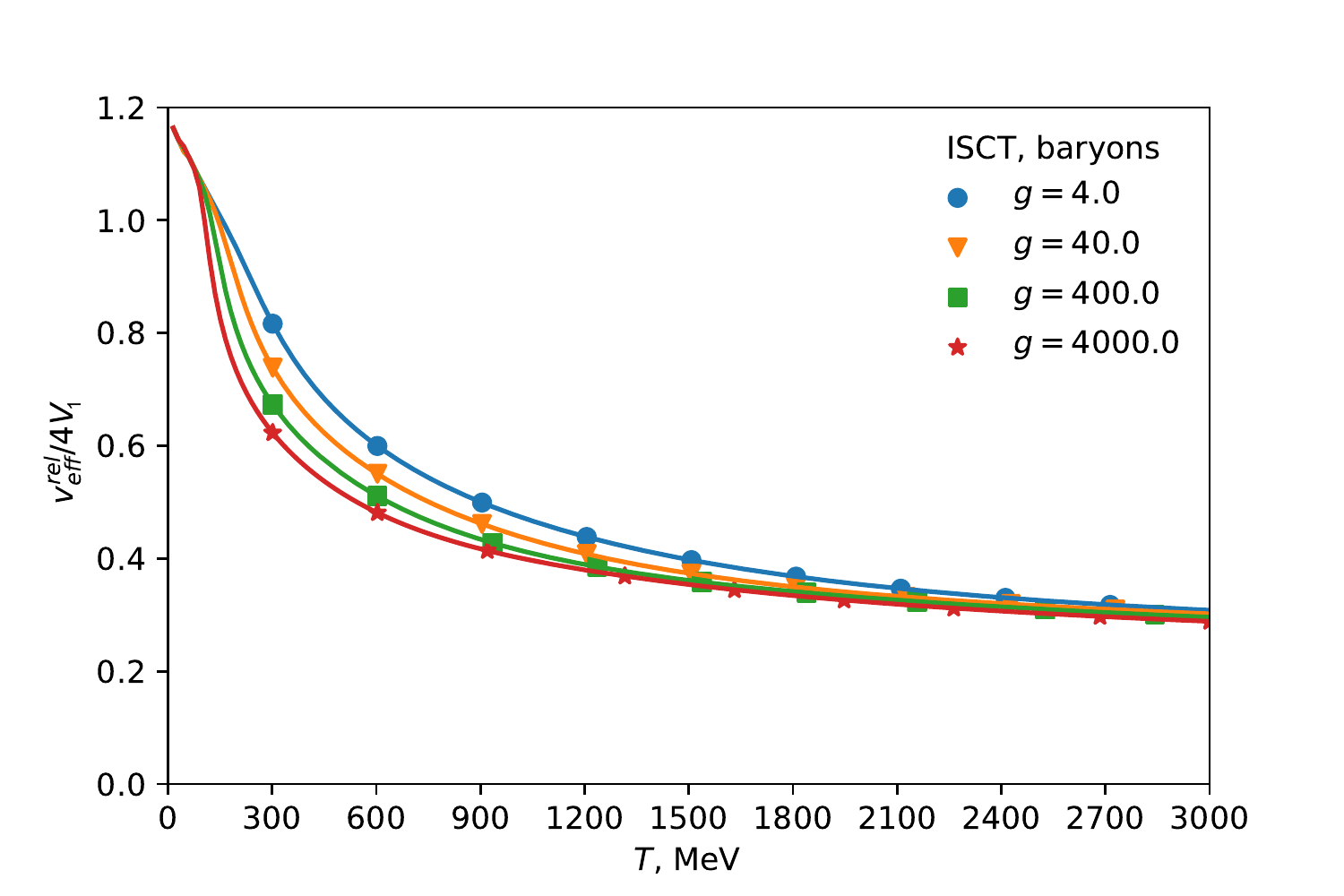}
	\includegraphics[width=0.5\textwidth]{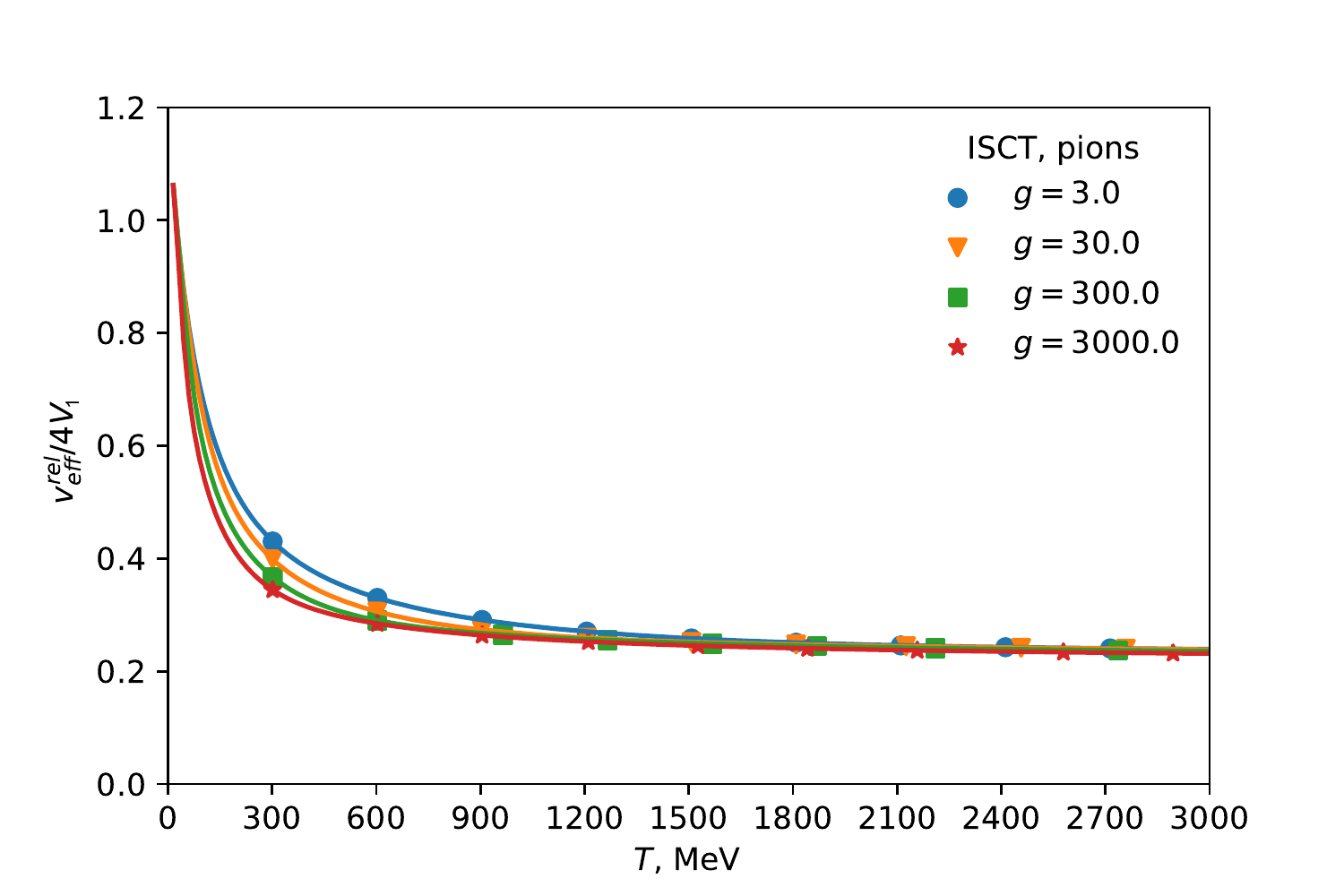}
	\caption{ Same as in figure \ref{fig:vol1} but for different degeneracy factors $g$. \textit{Left panel:} a mixture of nucleons and anti-nucleons. \textit{Right panel:} a mixture of pions.}
	\label{fig:vol2}
\end{figure}

Figure \ref{fig:vol1} shows the dependence of the averaged effective excluded volume $\langle v_1^{eff} \rangle $ (\ref{EqVIp}) for the gas of baryons (left panel) and the gas of pions (right panel) for different sets of ISCT EoS parameters while figure \ref{fig:vol2} shows the same for a wide range of degeneracy factors $g$ up to $4000$ and $3000$ respectively. These high $g$ values should be regarded as a necessary check-up to employ the developed EoS to model the properties of mixtures of all known hadrons. High values of temperature for which the hadrons cannot exist in nature are considered here to study the dense packing limit of the developed ISCT EoS.

From figures \ref{fig:vol1} and \ref{fig:vol2} one can see that at high temperatures the averaged effective excluded volume is $\langle v_1^{eff} \rangle \simeq \frac{4}{3} V_1$, i.e. it is larger than the eigenvolume of particles $V_1$. Hence, besides the eigenvolume of particle $V_1$ the averaged effective excluded volume $\langle v_1^{eff} \rangle $ acquires an additional contribution from the surface term as discussed in the preceding section. This result means that the maximal density of studied systems approaches the value $\frac{1}{ \langle v_1^{eff} \rangle} \simeq \frac{3}{4 V_1} = \frac{0.75}{V_1}$. In other words, the maximal packing fraction $\eta \simeq \frac{ V_1}{\langle v_1^{eff} \rangle} $ of studied systems automatically approaches a dense packing limit of equal spheres $\eta_{exc} = \pi/\left(3\sqrt{2}\right) \approx 0.74$ with sufficient accuracy and without any prior knowledge about such system configuration for all studied mixtures. Although, in contrast to pions, the gas of baryons is still approaching a dense packing limit, since the baryons are essentially heavier and, hence, not all of them are ultra-relativistic even at so high values of $T$, our analysis shows that at higher densities (and $T$ values) the baryons behave similarly to pions. In our opinion, these are remarkable and intriguing findings.

Moreover, these findings lead us to formulate a natural question that may be of principal importance for the statistical mechanics: do the relativistic rigid spheres with the Lorentz contraction, like their non-relativistic counterparts, have the gas to solid phase transition? The classical hard spheres demonstrate the existence of the gas to solid phase transition at the packing fractions $ 0.49< \eta <0.55$ \cite{Simple_Liquids, Mulero}, but the famous Carnahan-Starling EoS \cite{CSeos} does not describe it. Therefore, the fact that the ISCT EoS of relativistic rigid spheres does not exhibit the gas to solid phase transition may mean only that to answer the question above, at high packing fractions one needs even more sophisticated EoS. This question may be of some importance for the EoS of neutron stars \cite{EOSns}.

The above results depicted in the figures \ref{fig:vol1} and \ref{fig:vol2} allow us to deduce that the averaged effective excluded volume
of hadrons become incompressible at temperatures above $T = 1500$ MeV and, hence, one can expect the break of causality under such conditions. Moreover, the increase of the degeneracy factor by three orders of magnitude does not qualitatively change the behavior of averaged effective excluded volume of hadrons. This means that the suggested ISCT EoS of relativistic rigid spheres can be used to model the properties of hadronic matter of QCD. 

\section{Conclusions} \label{sec:concl}

The present work is a first and important step to the development of the novel hadron resonance gas model with multicomponent hard-core repulsion which is causal inside the whole hadronic phase of the QCD matter. In contrast to all known formulations of the hadronic matter EoS which in some way are taking into account the Lorentz contraction of particle's eigenvolumes, the present formulation correctly reproduces the relativistic excluded volume of two hard-core particles with arbitrary velocities and masses. This automatically provides the correct value of their second virial coefficient, which in the relativistic case differs from the relativistic excluded volume. 

This achievement was made possible by obtaining an analytical formula that rather accurately accounts for the Lorentz contraction of the excluded volume of two relativistic hadrons with hard-core repulsion. Moreover, based on this analytical formula we heuristically derived the induced surface and curvature tension EoS for Boltzmann particles with relativistic excluded volumes. Such an EoS is formulated in terms of system pressure $p$, its surface $\Sigma$ and curvature tension $K$ coefficients. In this way, we extend the morphological thermodynamics approach to the systems of Lorentz contracted rigid spheres, but in the grand canonical ensemble since in the relativistic statistical mechanics, the number of hadrons is not conserved. Also, we argue that the bending rigidity coefficient $\ \psi $ (related to the Euler characteristic $X_B$ of particle $B$), which exists in the standard formulation of morphological thermodynamics, is not necessary for the grand canonical formulation because its contribution to the particle-free energy just shifts the value of particle's chemical potential. Moreover, in contrast to particle eigenvolume $V_B$, eigensurface $S_B$ and doubled eigenperimeter $C_B$, the Euler characteristic $X_B$ simply does not appear in our equations. Therefore, we believe it can be omitted at this stage of our research.

In our analysis of effective excluded volumes of lightest baryons and mesons, we found some peculiar behavior at very high temperatures. Namely, independently of the employed parameterization, the mass of particles and the value of their degeneracy factors (we used degeneracy factors up to 3000 and 4000 units), the ISCT EoS leads to the same maximal packing fraction $\eta \simeq 0.75$ of Lorentz contracted particles which is very close to a dense packing limit of classical hard spheres of same radius $\eta_{exc} \approx 0.74$.

We believe this is an interesting finding since it leads us to a new problem for statistical mechanics of phase transitions of relativistic particles, namely what may happen to the gas-solid phase transition of classical hard spheres, if one takes into account their Lorentz contraction? Neither the EoS derived here, nor the one, developed in \cite{Bugaev_2} and proved to be causal at high pressure, show any traces of such phase transition for the Lorentz contracted rigid spheres. But it also may be that a more sophisticated analysis of this question can answer it rigorously. Note that this question may be not only of academic importance, but it may be important for the EoS of neutron matter which recently attracts a lot of attention. 

\ack
The authors are thankful to Boris Grinyuk, Oleksandr Vitiuk and Ivan Yakimenko for the valuable comments.
KAB acknowledges support from the NAS of Ukraine by its priority project ``Fundamental properties of the matter in the relativistic collisions of nuclei and in the early Universe" (No. 0120U100935).
DBB received funding from the Polish National Science Center (NCN) under grant No. 2019/33/B/ST9/03059 and from the Russian Fund for Basic Research under grant No. 18-02-40137.
The work of LVB, EEZ, NSYa and KAB was supported in part by the Norwegian Centre for International Cooperation and Quality Enhancement in Higher Education (DIKU) under grants CPEA-LT-2016/10094 and UTF-2016-long-term/10076. LVB and EEZ acknowledge the support of the Norwegian Research Council (NFR) under grant No. 255253/ F53 CERN Heavy Ion Theory and the Russian Foundation for Basic Research (RFBR) under the grants 18-02-40085 and 18-02-40084.
The authors are grateful to the COST Action CA15213 ``THOR" for supporting their networking. \\

\section*{References}
\bibliography{bibliography}

\begin{thebibliography}{10}
\expandafter\ifx\csname urlstyle\endcsname\relax
  \providecommand{\doi}[1]{doi:\discretionary{}{}{}#1}\else
  \providecommand{\doi}{doi:\discretionary{}{}{}\begingroup
  \urlstyle{rm}\Url}\fi

\bibitem{MorTer1}
P.-M. K\"onig, R.~Roth and K.~R. Mecke,
  \href{http://dx.doi.org/10.1103/PhysRevLett.93.160601}{{\em Phys. Rev. Lett.}
  {\bf 93}},   160601 (Oct 2004).

\bibitem{MorTer2}
H.~Hansen-Goos and R.~Roth,
  \href{http://dx.doi.org/10.1088/0953-8984/18/37/002}{{\em J. Phys.: Condens.
  Matter} {\bf 18}}, 8413 (Aug 2006).

\bibitem{MorTer3}
R.~Roth, K.~Mecke and M.~Oettel,
  \href{http://dx.doi.org/10.1063/1.3687921}{{\em J. Chem. Phys.} {\bf 136}},
  081101  (2012).

\bibitem{Hadwiger1957}
H.~Hadwiger, {\em Vorlesungen Über Inhalt, Oberfläche und Isoperimetrie}, 1st
  ed. edn. (Springer, Berlin, Germany, 1957).

\bibitem{HadwigerTheorem}
K.~R. Mecke, {\em Int. J. Mpd. Phys. B} {\bf 12}, 861  (1998).

\bibitem{Nazar19}
N.~S. Yakovenko, K.~A. Bugaev, L.~V. Bravina and E.~E. Zabrodin,
  \href{http://dx.doi.org/https://doi.org/10.1140/epjst/e2020-000036-3}{{\em
  Eur. Phys. J. Special Topics} {\bf 229}}, 3445  (2020).

\bibitem{QSTAT2019}
K.~A. Bugaev, \href{http://dx.doi.org/10.1140/epja/i2019-12920-2}{{\em Eur.
  Phys. J. A} {\bf 55}},   215  (2019).

\bibitem{LFtrans3}
K.~A. Bugaev, \href{http://dx.doi.org/10.1088/1361-6471/abce92}{{\em J. Phys.
  G} {\bf 48}},   055105 (Apr 2021).

\bibitem{Kapusta}
J.~I. Kapusta and K.~A. Olive,
  \href{http://dx.doi.org/https://doi.org/10.1016/0375-9474(83)90241-5}{{\em
  Nucl. Phys. A} {\bf 408}}, 478   (1983).

\bibitem{Zhang}
Q.-R. Zhang, \href{http://dx.doi.org/https://doi.org/10.1007/BF01292341}{{\em
  Z. Phys. A} {\bf 353}}, 345  (1995).

\bibitem{Bugaev_1}
K.~A. Bugaev, M.~I. Gorenstein, H.~St{\"o}cker and W.~Greiner,
  \href{http://dx.doi.org/https://doi.org/10.1016/S0370-2693(00)00690-0}{{\em
  Phys. Lett. B} {\bf 485}}, 121  (2000).

\bibitem{Bugaev_2}
K.~A. Bugaev,
  \href{http://dx.doi.org/https://doi.org/10.1016/j.nuclphysa.2008.04.007}{{\em
  Nucl. Phys. A} {\bf 807}}, 251  (2008).

\bibitem{Indians20}
S.~Pal, A.~Bhattacharyya and R.~Ray,
  \href{http://dx.doi.org/10.1016/j.nuclphysa.2021.122177}{{\em Nucl. Phys. A}
  {\bf 1010}},   122177  (2021).

\bibitem{Mayer}
J.~E. Mayer and M.~G. Mayer, {\em Statistical mechanics}, 2nd ed. edn. (Wiley
  New York, 1977).

\bibitem{IST3}
K.~A. Bugaev, V.~V. Sagun, A.~I. Ivanytskyi, I.~P. Yakimenko, E.~G. Nikonov,
  A.~V. Taranenko and G.~M. Zinovjev,
  \href{http://dx.doi.org/10.1016/j.nuclphysa.2017.11.008}{{\em Nucl. Phys. A}
  {\bf 970}}, 133  (2018).

\bibitem{X2}
V.~V. Sagun, K.~A. Bugaev, A.~I. Ivanytskyi, I.~P. Yakimenko, E.~G. Nikonov,
  A.~V. Taranenko, C.~Greiner, D.~B. Blaschke and G.~M. Zinovjev,
  \href{http://dx.doi.org/10.1140/epja/i2018-12535-1}{{\em Eur. Phys. J. A}
  {\bf 54}},   100 (Jun 2018).

\bibitem{X3}
K.~A. Bugaev {\em et~al.},
  \href{http://dx.doi.org/10.1140/epja/s10050-020-00296-5}{{\em Eur. Phys. J.
  A} {\bf 56}},   293  (2020).

\bibitem{X4}
O.~V. Vitiuk, K.~A. Bugaev, E.~S. Zherebtsova, D.~B. Blaschke, L.~V. Bravina,
  E.~E. Zabrodin and G.~M. Zinovjev,
  \href{http://dx.doi.org/10.1140/epja/s10050-021-00370-6}{{\em Eur. Phys. J.
  A} {\bf 57}},  ~74  (2021).

\bibitem{X5}
K.~A. Bugaev {\em et~al.}  (2021),
  \href{http://arxiv.org/abs/2104.05351}{{\ttfamily arXiv:2104.05351
  [hep-ph]}}.

\bibitem{PDG}
P.~A. Zyla {\em et~al.}, \href{http://dx.doi.org/10.1093/ptep/ptaa104}{{\em
  Prog. Theor. Exp. Phys.} {\bf 2020}} (08 2020), 083C01.

\bibitem{Sagun14}
V.~Sagun, A.~Ivanytskyi, K.~Bugaev and I.~Mishustin,
  \href{http://dx.doi.org/https://doi.org/10.1016/j.nuclphysa.2013.12.012}{{\em
  Nucl. Phys. A} {\bf 924}}, 24  (2014).

\bibitem{Roth1}
P.-M. K{\"o}nig, R.~Roth and K.~Mecke,
  \href{http://dx.doi.org/https://doi.org/10.1103/PhysRevLett.93.160601}{{\em
  Phys. Rev. Lett.} {\bf 93}},   160601  (2004).

\bibitem{Roth2}
P.-M. K{\"o}nig, P.~Bryk, K.~Mecke and R.~Roth,
  \href{http://dx.doi.org/https://doi.org/10.1209/epl/i2004-10410-4}{{\em
  Europhys. Lett.} {\bf 69}},   832  (2005).

\bibitem{Ivanytskyi18}
A.~Ivanytskyi, K.~Bugaev, V.~Sagun, L.~Bravina and E.~Zabrodin,
  \href{http://dx.doi.org/https://doi.org/10.1103/PhysRevC.97.064905}{{\em
  Phys. Rev. C} {\bf 97}},   064905  (2018).

\bibitem{CSeos}
N.~F. Carnahan and K.~E. Starling, {\em J. Chem. Phys.} {\bf 51}, 635  (1969).

\bibitem{Simple_Liquids}
J.-P. Hansen and I.~R. McDonald, {\em Theory of simple fluids} (Academic Press,
  Amsterdam, 2006).

\bibitem{Mulero}
A.~Mulero, Theory and simulation of hard-sphere fluids and related systems, in
  {\em 1st ed.\/}, , Lect. Notes Phys. Vol.~753 (Springer-Verlag Berlin
  Heidelberg, 2008) p. 546.

\bibitem{EOSns}
K.~A. Bugaev {\em et~al.},
  \href{http://dx.doi.org/10.3390/universe5020063}{{\em Universe} {\bf 5}},
  ~63  (2019).

\end{thebibliography}

\end{document}